%% file: main.tex
\documentclass[]{spie}

\input{Other/commands}

\usepackage[OT2,T1]{fontenc}
\DeclareSymbolFont{cyrletters}{OT2}{wncyr}{m}{n}
\DeclareMathSymbol{\Sha}{\mathalpha}{cyrletters}{"58}
 
\usepackage{amsmath,amsfonts,amssymb,esint}
\usepackage{graphicx}
\usepackage{lipsum}
\usepackage{natbib}
\usepackage[colorlinks=true, allcolors=blue]{hyperref}
\usepackage{multicol}
\usepackage{subcaption}
\usepackage{times}
\usepackage{setspace}
\usepackage{lineno}

\title{Episodic fluid venting from sedimentary basins fuelled by \\ pressurised mudstones}
\author[1]{Luke~M.~Kearney}
\author[1]{Richard~F.~Katz}
\author[2]{Christopher~W.~MacMinn}
\author[1]{Chris~Kirkham}
\author[1]{Joe~Cartwright}
\affil[1]{\normalsize{\textit{Department of Earth Sciences, University of Oxford, Oxford OX1 3AN, United Kingdom}}}
\affil[2]{\normalsize{\textit{Department of Engineering Science, University of Oxford, Oxford OX1 3PJ, United Kingdom}}}

\begin{document}

\maketitle

\vspace{0.5cm}

\noindent
\textbf{Subsurface sandstone reservoirs sealed by overlying, low-permeability layers provide capacity for long-term sequestration of anthropogenic waste. Leakage can occur if reservoir pressures rise sufficiently to fracture the seal. Such pressures can be generated within the reservoir by vigorous injection of waste or, over thousands of years, by natural processes. In either case, the precise role of intercalated mudstones in the long-term evolution of reservoir pressure remains unclear; these layers have variously been viewed as seals, as pressure sinks or as pressure sources. Here, we use the geological record of episodic fluid venting in the Levant Basin to provide striking evidence for the pressure-source hypothesis. We use a Bayesian framework to combine recently published venting data, which record critical subsurface pressures since $\sim$2~Ma, with a stochastic model of pressure evolution to infer a pressure-recharge rate of $\sim$30~MPa/Myr. To explain this large rate, we quantify and compare a range of candidate mechanisms. We find that poroelastic pressure diffusion from mudstones provides the most plausible explanation for these observations, amplifying the $\sim$3~MPa/Myr recharge caused primarily by tectonic compression. Since pressurised mudstones are ubiquitous in sedimentary basins, pressure diffusion from mudstones is likely to promote seal failure globally.}

Sedimentary successions often include high-permeability sandstone units enveloped by thick, low-permeability mudstone units. Because the surrounding mudstones can act as barriers to fluid leakage, these sandstones are often viewed as sealed reservoirs and therefore as targets for the large-scale sequestration of waste or storage of sustainable fuels \citep{krevor2023subsurface, heinemann2021enabling, ringrose2021storage}. However, fluid injection can pressurise such a reservoir to the point of triggering hydraulic fractures that breach the mudstone seal, enabling rapid depressurisation by fluid venting. This mechanism of sediment depressurisation has been recognised for several decades \citep{noble1963formation, cathles1983thermal, roberts1995episodic}. It is generally believed that pressures below this failure threshold will dissipate by poroelastic diffusion through sealing mudstones over thousands of years~\citep{muggeridge2004dissipation, muggeridge2005rate, chang2013dissipation, luo2016overpressure}. However, this slow depressurisation relies on the assumption that the mudstones themselves will remain at low pressure over these long timescales, whereas a variety of natural mechanisms are known to gradually pressurise the entire sedimentary column~\citep{osborne1997mechanisms}. Luo \& Vasseur \citep{luo2016overpressure} showed that overpressured mudstones can, in theory, act as a pressure source rather than as a pressure sink, re-pressurising a sandstone reservoir after natural fluid venting. They proposed that this mechanism could fuel further episodes of venting. \cite{kearney2023episodic} recently developed a poroelastic model of episodic venting that supports and extends this basic concept. However, the predictions of these theoretical studies are difficult to test against observational evidence due to the long timescale associated with mudstone pressure evolution.

\begin{figure*}[!t]
  \centering
  \includegraphics[width=1\linewidth]{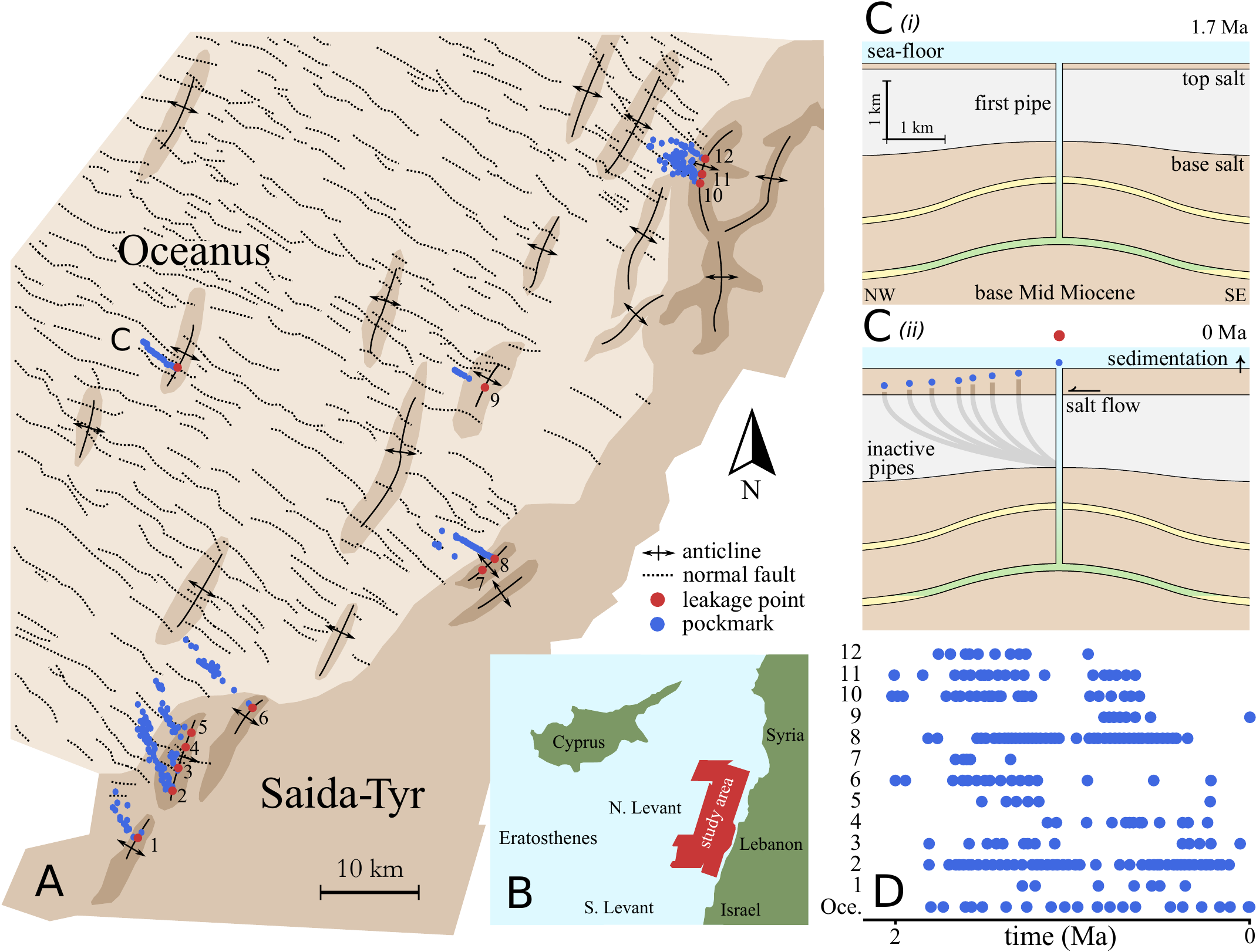}
\caption{Fluid escape pipe trails in the Levant Basin. (A) Overview of base-salt surface, showing sub-salt anticlines and the elevated margin platform, adjacent to the normally faulted deeper basin; adapted from \cite{oppo2021leaky}, where lighter colours indicate larger depth. (B) Study area located on the North Levant Basin margin, offshore Lebanon. (C) General mechanism for fluid escape pipe trail formation, adapted from \cite{cartwright2018direct}, with (\textit{i}) as the formation of the initial pipe at 1.7 Ma and (\textit{ii}) as the present-day arrangement. (D) Pipe trails labelled 1--12 and Oceanus from panel (A) when corrected for relative salt translation rates \citep{oppo2021leaky}.
\label{fig:1}}
\end{figure*}

Here, we test the hypothesis that mudstones can act as sources of pressure, fuelling fluid venting from sedimentary basins. The geological record of episodic fluid venting in the Levant Basin (Fig.~\ref{fig:1}A) provides a rare opportunity to elucidate the role of mudstones in the pressure evolution of sedimentary basins. These vents release overpressure in localised fluid-expulsion events that transport fluid through kilometres of low-permeability rock via cylindrical conduits known as fluid-escape pipes \citep{cartwright2018direct}. These pipes provide a high-permeability pathway to the surface, where they terminate as pockmarks, each recording a discrete episode of venting. Field observations of relict fluid-escape pipes consistently show evidence of fracturing \citep{huuse2005giant, roberts2010structure, loseth20111000}, suggesting that these pipes form by hydraulic fracturing \citep{cartwright2015seismic}. Hydraulic fracturing typically requires the pore pressure to exceed the local compressive stress; indeed, drilling in a region of active venting has revealed near-lithostatic pore pressures \citep{reilly2010deep}. Furthermore, the resulting pockmarks enable stratigraphic estimates of the time of each venting episode and thus constrain the rate of pressure recharge between episodes.

\section*{The Levant Basin}

In the North Levant Basin, located in the Eastern Mediterranean (Fig.~\ref{fig:1}B), more than 300 fluid-escape pipes have been documented, recording episodic fluid venting from 13 fixed locations across the region. For one of these locations, named Oceanus (Fig.~\ref{fig:1}C), \cite{cartwright2021quantitative} calculated that the initiation of venting via hydraulic fracturing requires $\sim$30~MPa of overpressure. Tectonic compression and marginal uplift have been proposed as the main overpressuring mechanisms in the region \citep{oppo2021leaky, cartwright2021quantitative}. The Levant Basin resides within a compressive tectonic regime stemming from the collision of the African and Eurasian plates. We estimate the strain at Oceanus to be less than 10\% (Supplementary Material~S1). Within the Levant Basin is a $\sim$3~km-thick Oligo-Miocene clastic succession consisting of turbiditic sandstones of Late Oligocene to Early Miocene age that are encased by mudstone. Many of these sandstone reservoirs host biogenic methane accumulations in NE--SW trending anticlines. The Levant pipes source methane and water from these anticline reservoirs and terminate at the seafloor as pockmarks (Fig.~\ref{fig:1}A). The pipes penetrate through a $\sim$1.5~km-thick layer of salt deposited during the Messinian Salinity Crisis \citep{ryan2009decoding}. Recent activity of the Levant Fracture System has been uplifting the eastern margin of the basin, leading to gravity-driven, basinward salt flow since $\sim$2~Ma, contemporaneous with the formation of pipes in the area. 

Each pipe forms vertically but the basinward viscous flow of salt advects existing pipes away from their initial positions, such that subsequent venting from the same reservoir requires the formation of a replacement pipe (Fig.~\ref{fig:1}C). Repetition of this process leads to the 13 observed trails of pipes in the North Levant Basin \citep{cartwright2018direct, oppo2021leaky}. Thus, each pipe trail records episodic fluid venting from a single reservoir, suggesting that these reservoirs are repeatedly repressurised. From the spatial distributions of pockmarks within each pipe trail, the time of formation of each pipe can be estimated (Fig.~\ref{fig:1}D) using the methods of \cite{oppo2021leaky} and \cite{cartwright2021quantitative}. These approaches reveal that for each trail, pipe formation typically occurs every $\sim$100~kyr. Since fluid-escape pipes record critical subsurface pressures, the Levant pipe trails enable us to distinguish between theories for pressure redistribution between sedimentary layers. The timings of the pockmarks of the isolated Oceanus pipe trail are particularly well constrained as it is situated in a less tectonically-active region of the basin. Oceanus is therefore less susceptible to local stress changes that might affect the recharge mechanics. We thus focus our analysis on the Oceanus trail. The remaining 12 trails are distributed along the active basin margin and are used to extend our inferences from Oceanus to a more complex system.

To test the pressure-source hypothesis, we develop a novel stochastic model of reservoir pressure evolution and use it to invert the Levant pipe trail data under a Bayesian framework for model parameters such as the pressure-recharge rate. Using basic physical arguments, we then estimate recharge rates for each candidate overpressure mechanism and compare with the inferred rates. In particular, \cite{kearney2023episodic} showed that pressure diffusion from mudstones amplifies the rate of pressure recharge generated by tectonic compression. In mudstone-dominated basins like the Levant Basin, pressure-recharge rates can be amplified by a factor of $\sim$10. Therefore if this hypothesis is correct, then we expect that the inferred recharge rate is a factor of $\sim$10 greater than that predicted for tectonic compression alone. 

\section*{Stochastic Model of Pressure Evolution}

We assume that a fluid-escape pipe forms via hydraulic fracturing when the pore pressure exceeds the critical fracture pressure $p_f$, which is the sum of the minimum horizontal compressive stress $\sigma_{\mathrm{min}}$ and tensile strength $\sigma_{T}$ of the overlying mudstone \citep{price1990analysis, scandella2011conduit},
\begin{equation}
    p_f = \sigma_{\mathrm{min}} + \sigma_T,
\end{equation}
where we take compression to be positive. Once venting begins, the pressure drops rapidly until the pathway closes, which we assume occurs when the pressure reaches $\sigma_{\mathrm{min}}$. Once closed, we expect fractures to self-seal via swelling and mineral precipitation \citep{bock2010self}. Roberts \& Nunn \citep{roberts1995episodic} predict fluid venting durations of order years, which may be considered instantaneous relative to recharge times, of order 100~kyrs. Over the latter timescale, pressure will become spatially uniform within a high-permeability reservoir. We thus assume that reservoir pressure depends on time only. For multiple venting episodes to be sourced from the same reservoir, the reservoir pressure must recharge between episodes. We consider generic pressure recharge at an average rate $\Gamma$, such that the corresponding time $\Delta t$ between events is
\begin{equation}
\label{eqn:dt}
    \Delta t = \frac{p_f - \sigma_{\mathrm{min}}}{\Gamma} = \frac{\sigma_T}{\Gamma}.
\end{equation}
Fractures exploit pre-existing rock weaknesses that change over geologic time, such that $\sigma_T$ will vary between events. We model this variability by asserting that $\sigma_T$ is a normally distributed random variable with mean $\overline{\sigma_T}$ and standard deviation $s_T$. Equation (\ref{eqn:dt}) then implies that $\Delta t \sim \mathcal{N}(\overline{\sigma_T}/\Gamma, s_T/\Gamma)$. Thus, the mean and standard deviation of inter-event times of a trail of pockmarks can be used to infer the underlying recharge rate. 

As this is a limited dataset that has been produced by an inherently stochastic process, Bayesian inference is used to invert the pipe trail data for the full probability distribution of each parameter and quantify their uncertainty. Our prior estimates of each parameter ($\Gamma, \overline{\sigma_T}, s_T$) are updated by evaluating the data with a likelihood function to recover the posterior probability distributions of each parameter. The likelihood function provides a statistical measure of model--data agreement by calculating the probability of observing the data given a set of model parameters. The simplicity of our physical model enables the likelihood function to be expressed analytically (Supplementary Material~S2). We apply a conservative Gaussian prior for $\overline{\sigma_T}~\sim~\mathcal{N}(2.0, 1.0)$~MPa, since mudstone tensile strengths are typically a few~MPa \citep{okland2002importance, raaen2006improved}; in particular, Roberts \& Nunn \citep{roberts1995episodic} predict a pressure drop of $\sim$2~MPa from venting. The posterior distributions of each parameter are sampled using the Metropolis-Hastings algorithm \citep{hastings1970monte}.

\begin{figure*}[!t]
  \centering
  \includegraphics[width=1\linewidth]{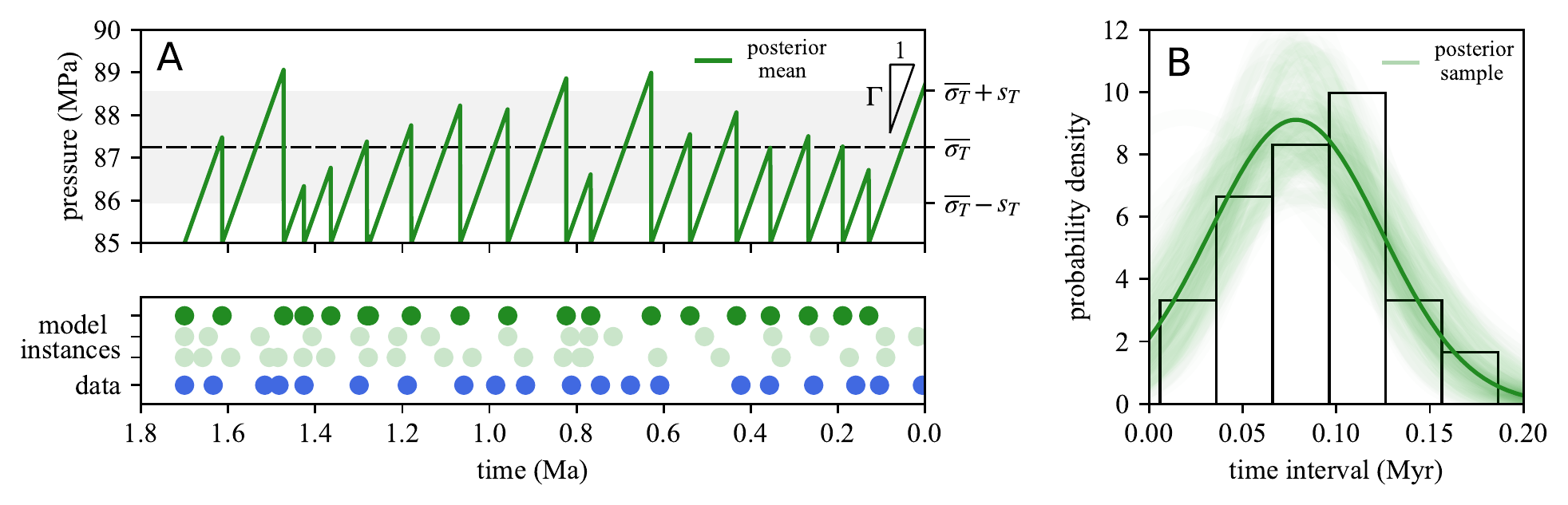}
\caption{Results of Bayesian inference applied to the Oceanus pockmark trail. (a) Lower panel: Time-transformed pockmark data (blue) and stochastic model instances of venting history using inferred posterior mean (dark green) and sample parameters (translucent green). Upper panel: Stochastic instance of linearised pressure evolution using inferred posterior mean parameters (dark green), with pressures in excess of the minimum compressive stress corresponding to the mean tensile strength (dashed line) and corresponding to tensile strength values within one standard deviation of the mean (grey shaded area). (b) Posterior mean (dark green) and sample (translucent green) time interval distributions compared with Oceanus data.
\label{fig:2} }
\end{figure*}

\section*{Oceanus Pockmark Trail}

From the posterior distributions inferred for the isolated Oceanus trail (Supplementary Material~S4), we use the mean posterior parameter values as input for our stochastic model to simulate an instance of linearised pressure evolution (Fig.~\ref{fig:2}A). The qualitative similarity between the pockmark data and the model output is apparent (Fig.~\ref{fig:2}A, lower panel). For statistical comparison, we use samples from the posterior parameter distributions to calculate a range of posterior time interval distributions (Fig.~\ref{fig:2}B) that agree well with the data; variations between samples indicate the level of uncertainty in the inference. We note that as we have inferred the time-averaged recharge rate, this linearised pressure evolution resembles the sawtooth behaviour that is predicted for recharge from tectonic compression only \citep{cartwright2021quantitative, kearney2023episodic}. However, our statistical model makes no physical assumptions regarding the mechanism or dynamics of pressure recharge between venting episodes. 

\section*{Levant Margin Pockmark Trails}

To the east of Oceanus are 12 other trails distributed along the basin margin \citep{oppo2021leaky}. Some of these trails originate from the same anticline, separated only by $\sim$1~km, and thus may be in hydraulic communication. To account for this possibility, we introduce pressure coupling as a feature of the model. For a coupled system of pipes, after any one pipe vents, the pressures of all pipes coupled to it reset to $\sigma_{\mathrm{min}}$ and a new $\sigma_T$ is sampled for each. Therefore, the pipe that vents pressure temporarily inhibits any coupled pipes from venting. Consequently, the pipes in a coupled system form complementary pockmark series (Supplementary Material~S5). If a group of pipes are instead uncoupled, each pipe behaves independently. This contrast between independent and complementary venting behaviour is a qualitative diagnostic for pressure coupling.

\begin{figure*}[!t]
  \centering
  \includegraphics[width=0.99\linewidth]{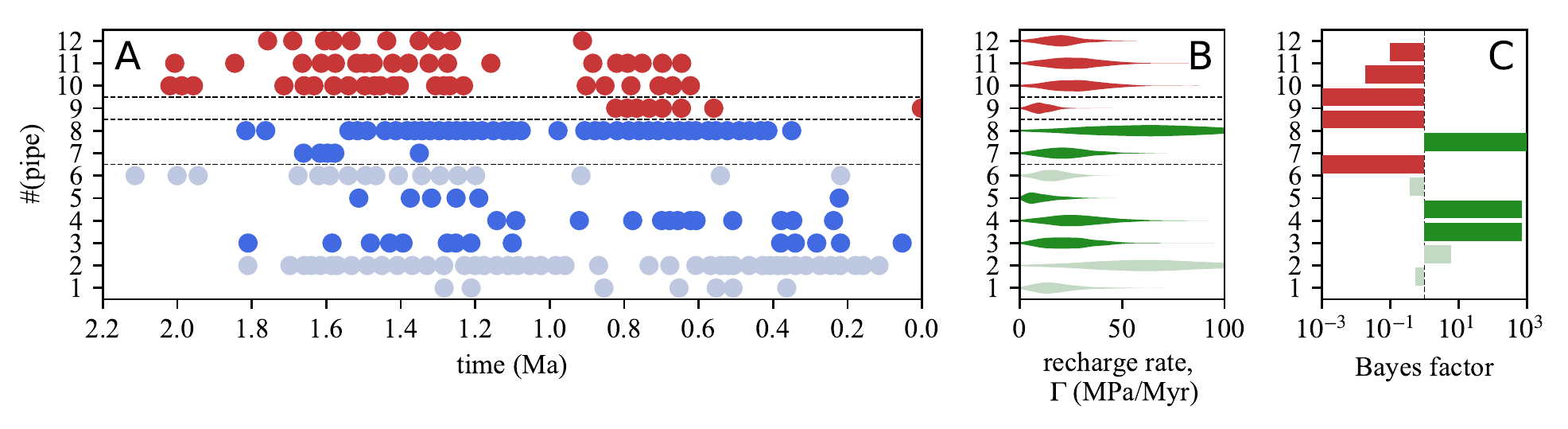}
\caption{Results of Bayesian inference applied to Levant margin data. (a) Time-transformed data from \cite{oppo2021leaky}. Dashed lines divide pipe clusters that are separated by more than 10 km. (b) Violin plot of posterior recharge rate distributions for each pipe trail. (c) Bayes factors of pairwise pipe analysis, where a positive value implies the coupled model is more likely. \label{fig:3} }
\end{figure*}

To evaluate whether a pair of adjacent trails are coupled, we calculate the Bayes factor of the coupled model $M_c$ and uncoupled model $M_u$ (Supplementary Material~S2). The Bayes factor $B_{cu}$ of two models $M_c$ and $M_u$ is given by the ratio of probabilities of observing the data $\textbf{t}$ given each model, i.e.,
\begin{equation}
    B_{cu} = \frac{\mathbb{P}(\textbf{t} \, | \, M_c)}{\mathbb{P}(\textbf{t} \, | \, M_u)}.
\end{equation}
For example, if $B_{cu} > 1$ then $M_c$ is preferred over $M_u$. Kass \& Raftery \citep{kass1995bayes} state that Bayes factors in the range 10--100 are `strong' and above 100 are `decisive'. We use this interpretation to assess the couplings of the pipe trails.

For the Levant margin pipe data (Fig.~\ref{fig:3}A), we infer similar recharge rates to those inferred for Oceanus, although mean recharge rates range up to 66~MPa/Myr for pipe trail 8 (Fig.~\ref{fig:3}B). Fig.~\ref{fig:3}C shows Bayes factors of pairwise analysis of adjacent trails. Triple-wise analysis leads to the same conclusions, but has been omitted to simplify the interpretation (Supplementary Material~S6). The model identifies all adjacent pipes that are greater than 10~km apart as decisively uncoupled. Furthermore, the inverted model indicates hydraulic connectivity between pipes 3, 4 and 5, each located along the same anticline, as well as trails 7 and 8 (Fig.~\ref{fig:3}C).

The inferences for pressure coupling are in agreement with the qualitative diagnostic behaviour. For example, the complementary venting behaviour of trails 3, 4 and 5 is visually evident. Conversely, trails 10, 11 and 12 are statistically inferred to be uncoupled and exhibit independent venting behaviour. Bayes factors with magnitudes below 10 exist for trail pairs $\{$1, 2$\}$, $\{$2, 3$\}$ and $\{$5, 6$\}$, indicating a lack of preference for either coupling or not. We attribute this neutrality to features in the data that obscure the underlying recharge mechanics. These features are likely due to local stress variations caused by, for example, faulting. Nonetheless, since the majority of results do have strong preferences to one model or another, we assert that the physical model captures the main pressure behaviour, both spatially and temporally. This result lends support to our statistical inferences of pressure-recharge rates. 

\section*{Comparison of Candidate Overpressure Mechanisms}

\begin{figure*}[!t]
  \centering
  \includegraphics[width=0.99\linewidth]{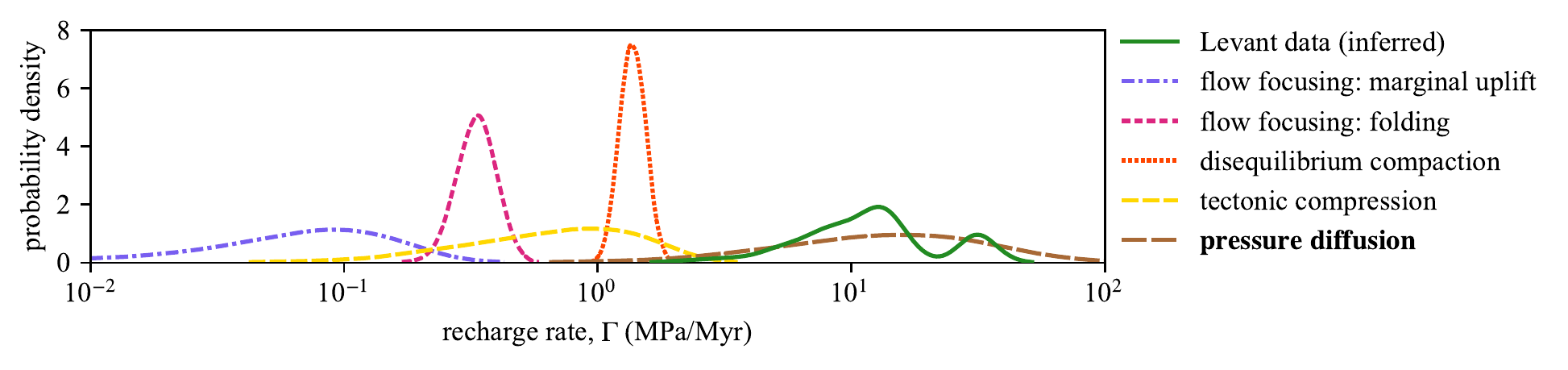}
\caption{Comparison of pressure-recharge rates inferred from Levant pipe trail data with estimated recharge rates from candidate mechanisms. \label{fig:4} }
\end{figure*}

The venting observations could plausibly be explained by various overpressure mechanisms that have been previously proposed. We next show that these mechanisms are inconsistent with our inferred recharge rates. Tectonic compression has been proposed as a major contributor to overpressure in the region \citep{cartwright2021quantitative}. Previous numerical modelling of tectonic compression indicates that overpressures of 11--14~MPa in total can be generated from 10\% strain \citep{obradors2016stress}. At Oceanus, the strain accumulated since at least the Messinian Salinity Crisis, 5--6~Ma, is less than 10\%. This implies a maximum recharge rate of $\sim$3~MPa/Myr from tectonic compression, which is insufficient to reproduce the observations (Fig.~\ref{fig:4}). However, \cite{kearney2023episodic} showed that pressure diffusion from mudstones amplifies the tectonic pressure-recharge rate in adjacent sandstones by a factor of $(1 + \nu/\gamma)$. The factor is termed the venting frequency multiplier and $\nu/\gamma$ is a ratio of dimensionless numbers that quantifies the relative effects of diffusion and compression. The dimensionless quantity $\gamma$ measures the tectonic pressure-recharge rate of the sandstone relative to that of the mudstone; $\nu$ is hydraulic capacitance of the mudstone relative to that of the sandstone. The hydraulic capacitance of a layer is the product of compressibility and thickness. Typically, $\nu/\gamma \gg 1$ in basins composed primarily of mudstone \citep{kearney2023episodic}, like the North Levant Basin. Due to the wide range of uncertainty in mudstone permeabilities \citep{yang2007permeability, chang2013dissipation}, it might be expected that the uncertainty in the recharge rate from mudstone pressure diffusion would span many orders of magnitude. However, the venting frequency multiplier is independent of the mudstone permeability \citep{kearney2023episodic}. This result enables us to calculate the recharge rate from the combined effect of diffusion and compression using prior distributions of each constituent parameter, giving a probability distribution that largely overlaps with inferred recharge rates (Fig.~\ref{fig:4}). 

Other candidate mechanisms predict much lower recharge rates than those inferred from the data (Fig.~\ref{fig:4}). The details of how we estimate the pressure-recharge rates from each mechanism are found in Supplementary Material~S7. \cite{oppo2021leaky} proposed that marginal uplift generates significant overpressures at the basin margin by driving lateral fluid migration from the highly overpressured deep basin. If pressure is transferred laterally along a connected, high-permeability sandstone unit, the venting periods would be several orders of magnitude lower than are observed. However, it is likely that there is poor lateral reservoir connectivity in the area \citep{cartwright2021quantitative} and our analysis above supports this idea, indicating that many relatively nearby pipes are likely to be hydraulically independent (Fig.~\ref{fig:3}C). As a result, the only pathway for lateral fluid migration is via mudstones, thus implicitly requiring pressure diffusion from mudstones for reservoir recharge. Marginal uplift may also generate overpressure by flow focusing \citep{flemings2002flow}, though this mechanism likely produces insufficient recharge rates (Fig.~\ref{fig:4}). Flow focusing due to fold amplification \citep{flemings2002flow} merely generates overpressures at a rate less than $\sim$1~MPa/Myr. Furthermore, hydrocarbon generation likely cannot generate the required recharge rate since the additional head required from buoyancy is greater than $\sim$1~km/Myr and most thermogenic gas generation was likely complete by 5--6~Ma \citep{al2016impact}. We cannot rule out the possibility of weak pressure recharge from biogenic gas generation, though petroleum systems modelling of the region favours biogenic gas accumulation via lateral migration from the deep basin \citep{bou20163d, ghalayini2018petroleum, nader2018key}. However, due to poor lateral reservoir connectivity, lateral gas migration is rate-limited by pressure diffusion (as for the case of marginal uplift). While lateral transfer produces insufficient recharge rates, vertical pressure transfer from a deeper reservoir along faults or fractures has been associated with fluid venting in other regions \citep[see][]{grauls1994role, tingay2007vertically, cathles2019processes}. In the Levant basin, however, there is no evidence to support vertical fluid migration. Moreover, vertical transfer cannot explain the observed regular periodicity of venting. Disequilibrium compaction due to the small, post-salt sediment accumulation of $\sim$300~m \citep{cartwright2021quantitative} creates a negligible pressurisation rate of $\sim$1~MPa/Myr. Sea-level fluctuations may trigger venting episodes \citep[see][]{scandella2011conduit}, though this mechanism alone provides no net pressure recharge. 

The venting observations from the Levant fluid-escape pipe trails are consistent with predictions deriving from the hypothesis that pressure diffusion from mudstones fuels episodic venting in the region. Therefore, the Levant pipe trails provide strong spatiotemporal evidence supporting this hypothesis. In doing so, the pipe trails support a more general idea---that pressure diffusion from mudstones plays an important role in pressure redistribution between sedimentary layers---and provide observational evidence that was previously lacking from the theoretical literature \citep[e.g.,][]{muggeridge2004dissipation, muggeridge2005rate, luo2016overpressure, kearney2023episodic}. It is likely that tectonic compression and marginal uplift were the main mechanisms for slowly pressurising the basin to near-lithostatic by $\sim$2~Ma. This pressurisation initiated basin-wide fluid venting by hydraulic fracturing, sourced by high-permeability, pre-salt sandstone reservoirs. Tectonic compression continued to slowly pressurise ($\sim$3~MPa/Myr) the entire sedimentary succession while poroelastic pressure diffusion from mudstones recharged the sandstone reservoirs back to failure at a rate of $\sim$30~MPa/Myr. This combination of pressure diffusion and tectonic compression, with minor contributions from hydrocarbon generation and disequilibrium compaction, led to episodic fluid venting with a typical venting period of $\sim$100~kyr. While this is not a universal result for pipes in any basin, pressure diffusion exists wherever the corresponding reservoir unit is encased by highly overpressured, low-permeability rocks. Furthermore, the effect of pressure diffusion is intensified in sedimentary basins composed mostly of mudstone \citep{kearney2023episodic}, where fluid venting phenomena are commonly observed \citep{cartwright2015seismic}. In many cases, liquefied mudstone is vented in addition to basinal fluids \citep[e.g.,][]{cartwright2023evolution}. The diverse roles of mudstones in pressure-driven, focused fluid venting provides an impetus to improve our mechanistic models of such venting phenomena.

\section*{Broader Implications}

Because understanding subsurface pressure is crucial to prevent unwanted fluid leakage, these results have wider implications for risk assessment during borehole drilling and the sequestration of waste such as CO$_2$. Fluid leakage resulting from reservoir pressurisation by mudstones may be a risk in a broad range of geological settings, requiring only that the mudstones are overpressured relative to the reservoir. This overpressure can be retained even after several episodes of fluid venting \citep{kearney2023episodic}, and can be generated by various means, not limited to horizontal compression. Indeed, \cite{kearney2023episodic} show that disequilibrium compaction (i.e., vertical compression) leads to mathematically equivalent behaviour. Therefore, even tectonically inactive regions like passive margins are prone to episodic venting if they are subjected to, for example, high sedimentation rates. Indeed, fluid-escape pipes are commonly observed in passive margin settings \citep{cartwright2015seismic}. Passive margins also provide the largest and likely most cost-effective large-scale CO$_2$ storage resource \citep{ringrose2019maturing}. Therefore, fluid-escape pipes may pose a significant threat to offshore storage projects.

This work highlights the importance of considering pressure diffusion from mudstones when assessing reservoir overpressures. This is especially true for sequestration sites with evidence of previous fluid venting, like the Sleipner field \citep{arts2004seismic, cavanagh2014sleipner}. While the relict fluid-escape pipes at Sleipner are unlikely to be a result of CO$_2$ injection \citep{cavanagh2014sleipner}, they serve as an example of the risks to containment associated with fluid venting. Although the dissolution of injected CO$_2$ can act to depressurise a storage reservoir \citep{akhbari2017causes}, evidence from a natural CO$_2$ reservoir suggests that the rate of depressurisation from CO$_2$ dissolution is $\sim$1~MPa/Myr \citep{sathaye2014constraints}. This is much less than the recharge rates from pressure diffusion that we infer in the Levant Basin, suggesting that CO$_2$ dissolution is unlikely to prevent leakage in regions where pressure diffusion from mudstones is active. Thus, for storage projects in regions with pressurised mudstones, our results indicate that reservoir pressure monitoring over several millenia may be required to ensure containment.

\bibliographystyle{agsm}
\footnotesize{\bibliography{main.bib}}

\setcounter{section}{0}
\renewcommand{\thesection}{S\arabic{section}}   
\renewcommand{\thefigure}{S\arabic{figure}}
\setcounter{figure}{0}
\renewcommand{\thetable}{S\arabic{table}}
\setcounter{table}{0}

\newpage
\section*{\titlefont Supplementary material}
\normalsize
\vspace{5mm}
\section{Oceanus strain}
\label{sec:strain}

\begin{figure*}[!htbp]
  \centering
  \includegraphics[width=1.0\linewidth]{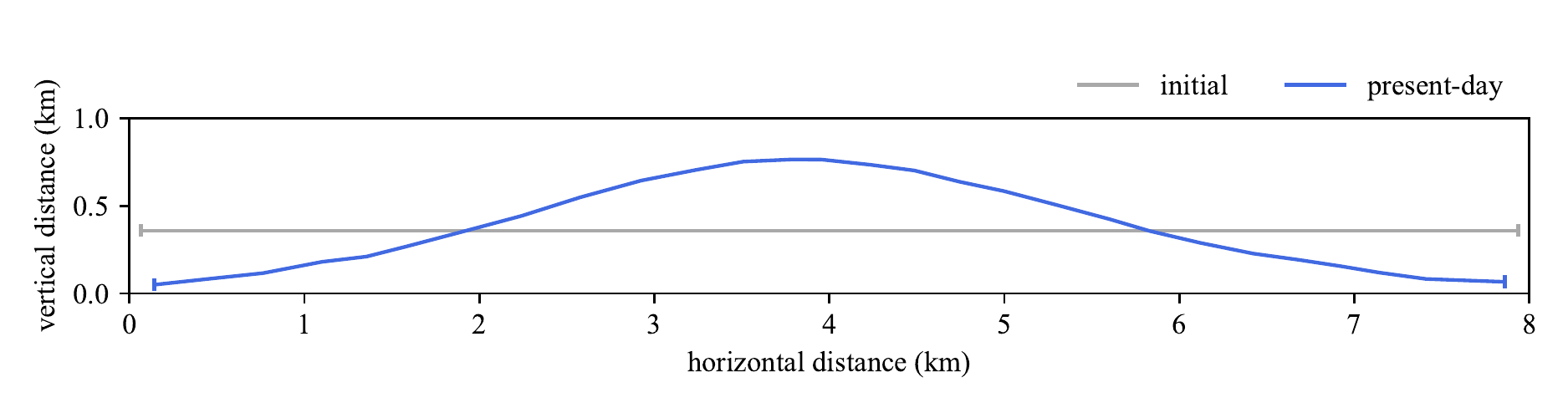}
\caption{Comparison of the present-day geometry of the folded sandstone layer at Oceanus (blue curve) with its likely initial state (grey curve), i.e.,~before the onset of tectonic compression. The horizontal strain at Oceanus can be calculated by comparing the horizontal extent of the present-day fold $L$ with the horizontal extent of its initial unfolded state $L_0$. \label{fig:strain_oceanus}}
\end{figure*}

We estimate the horizontal strain at the Oceanus pipe trail from the depth-converted cross-section in \cite{cartwright2021quantitative}. We do this by measuring the arc length of one of the folded Mid-Miocene sandstone layers (Fig.~\ref{fig:strain_oceanus}). We assume that this layer was initially horizontal, such that the arc length of the layer measures the initial horizontal extent of this section $L_0$. We calculate the horizontal strain by comparing this initial horizontal extent to the present-day horizontal extent of this layer $L$, using
\begin{equation}
    e_{xx} = -\frac{L - L_0}{L_0},
\end{equation}
where compressive strains are taken to be positive. This gives a horizontal strain of 2\% at Oceanus. While the majority of strain in the region is accommodated by folding, we recognise that this calculation does not account for strain accommodated by faulting or volumetric compression. Further uncertainty in this calculation stems from potential errors in the depth migration of the seismic data. Accounting for this uncertainty conservatively, we assert that the maximum possible horizontal strain at Oceanus is 10\%.

\section{Bayesian inference}
\label{sec:bayes}

One can use a Bayesian framework to invert for the parameters. Bayes' theorem is given by
\begin{equation}
    \prob(\theta \given \mathbf{x} ) = \frac{\prob(\mathbf{x} \given \theta ) \prob(\theta )}{\prob(\mathbf{x})} = \frac{\prob(\mathbf{x} \given \theta ) \prob(\theta )}{ \int_{\Theta} \prob(\mathbf{x} \given \theta ) \prob(\theta) \infd \theta },
\end{equation}
or in words,
\begin{equation}
    \mathrm{posterior} = \frac{\mathrm{likelihood} \times \mathrm{prior}}{\mathrm{evidence}}.
\end{equation}
Maximum-likelihood estimation methods such as ordinary least-squares aim to maximise the likelihood, the probability that the model generates the data. This is equivalent to maximising the posterior under the assumption of a uniformly distributed prior. However, Bayesian methods place a prior distribution on the parameters and calculates the posterior distribution using the observed data. We achieve the statistical equivalent to regularisation by enforcing these prior distributions. 

\subsection{Likelihood function}

Given a model, the likelihood function is the joint probability of the observed data. Here, the observed data is the set of all venting times~$\mathbf{t} = \{t_n\}_{n=1}^N$. The likelihood function can be decomposed in the following way:
\begin{equation}
    f(\{t_n\}_{n=1}^N) = \prod_{n=1}^N f(t_n \given \history),
\end{equation}
where $f$ is the probability density and the history $\history$ is the set of all event times until (but not including) $t_n$. Since the proposed model asserts that the pressure resets to $\sigma_{\mathrm{min}}$ after each event, the pressure `memory' of the system extends only from the most recent event so $\history = t_{n-1}$. We can therefore write
\begin{equation}
    f(\{t_n\}_{n=1}^N) = \prod_{n=1}^N f(\Delta t_n),
\end{equation}
where $\Delta t_n = t_n - t_{n-1}$. For coupled systems we must additionally consider the mark of each pipe $\kappa$, denoting where each event originated. It can be shown that for a set of $K$ coupled pipes,
\begin{equation}
    f(\{t_n, \kappa_n\}_{n=1}^N) = \prod_{n=1}^N f_{\kappa_n}(\Delta t_n) \prod_{k=1}^K \Big[ 1 - F_k(\Delta t_n) \Big]^{1 - \delta{\kappa_n, k}},
\end{equation}
where $f_k$ and $F_k$ are the uncoupled probability and cumulative density functions of pipe $\kappa = k$, respectively, and $\delta$ is the Kronecker delta. We utilise these likelihood functions to model the probability density of any coupling configuration of pipes. 

\subsection{Bayes factor}

To evaluate whether a pair of adjacent trails are coupled, we calculate the Bayes factor of the coupled model $M_c$ and the uncoupled model $M_u$. The Bayes factor $B_{cu}$ of two models $M_c$ and $M_u$ is given by the ratio of probabilities of observing the data $\textbf{t}$ given each model, i.e,
\begin{equation}
    B_{cu} = \frac{\prob(\textbf{t} \given M_c)}{\prob(\textbf{t} \given M_u)}.
\end{equation}
For example, if $B_{cu} > 1$ then $M_c$ is preferred over $M_u$. Here, $M_c$ is the coupled model and $M_u$ is the uncoupled model. \cite{kass1995bayes} state that Bayes factor magnitudes between 10-100 are `strong' and above 100 are `decisive'. We define a new parameter $\phi \in \{0, 1\}$ such that $\phi = 1$ indicates the coupled model $M_c$ and $\phi = 0$ indicates the uncoupled model $M_u$. The Bayes factor can be rewritten in terms of $\phi$ as 
\begin{equation}
    B_{cu} = \frac{\prob(\textbf{t} \given \phi = 1)}{\prob(\textbf{t} \given \phi = 0)}.
\end{equation}
In this form, the Bayes factor can be calculated with MCMC methods. We assume a prior distribution for $\phi \sim \textrm{Bernoulli}(\tfrac{1}{2})$. Using the likelihood functions of the coupled and uncoupled models, the posterior distribution $\prob(\mathbf{t} \given \phi)$ can be sampled (using e.g., the Metropolis-Hastings algorithm) from which the Bayes factor can be calculated.

\section{Inversion tests}

Inversions are performed on synthetic data to test the accuracy and sensitivity of our inversion method. Here, we apply uniform prior distributions to assess the effectiveness of the likelihood function alone. We nondimensionalise the parameters ($\Gamma, \overline{\sigma_T}, s_T$) to $\Gamma^* = \Gamma/\overline{\sigma_T}$ and $s_T^* = s_T/\overline{\sigma_T}$. 

\subsection{One pipe}

Inversions of a single pipe trail perform well if the true $s_T^* < 1$. When $s_T^* > 1$, the distribution of $\Delta t$ begins to be significantly truncated for $\Delta t < 0$ and tends towards a uniform distribution for larger $s_T^*$. Henceforth, we analyse simulations with $s_T^* < 1$.
\begin{figure}[!htbp]
\centering
\includegraphics[width=1.0\linewidth]{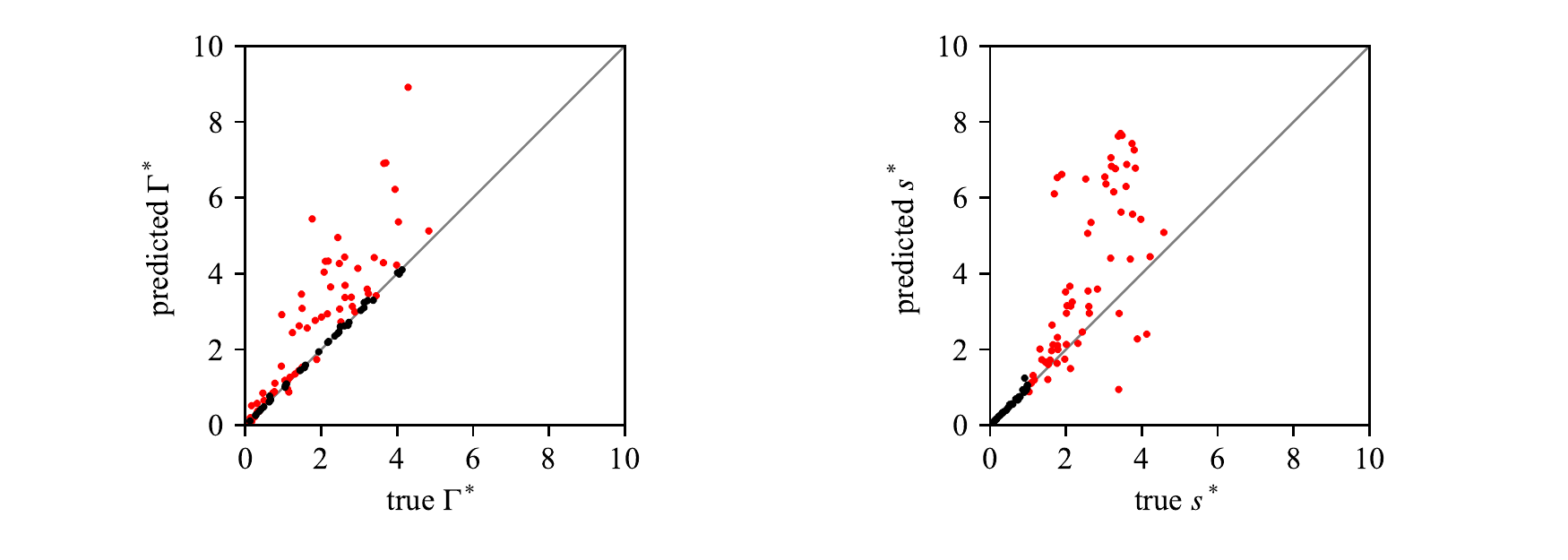}
\caption{Showing inversion results for simulated synthetic data. Each simulation uses a pair of values for $\Gamma^*$ and $s_T^*$ to generate a sequence of 1000 venting times. (a) Predicted mean $\Gamma^*$ versus the true assigned $\Gamma^*$ for that simulation. (b) Predicted mean $s_T^*$ versus the true assigned $s_T^*$ for that simulation. Black points represent inversion results from simulations with true $s^* < 1$ and red points with true $s^* > 1$.
\label{fig:inv_one}}
\end{figure}

\subsection{Two pipes}

\begin{figure}[!htbp]
\centering
\includegraphics[width=1.0\linewidth]{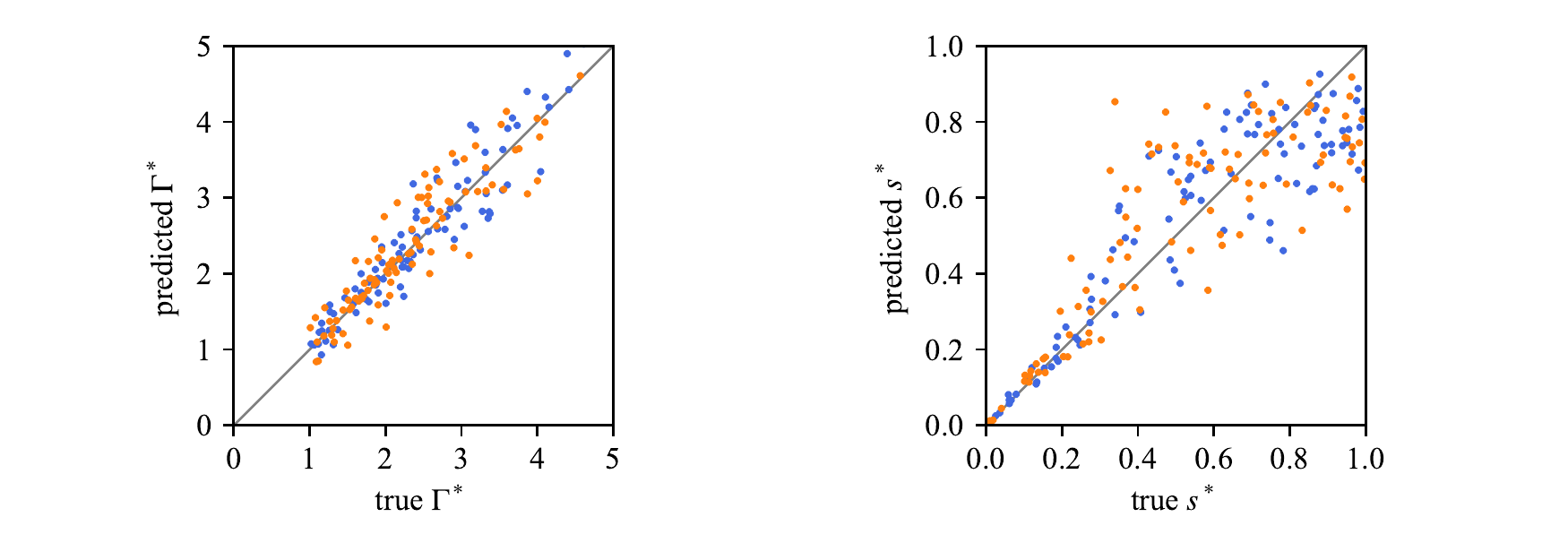}
\caption{Showing inversion results for simulated synthetic data of two uncoupled pipes. Each simulation uses a pair of values for $\Gamma^*$ and $s_T^*$ to generate 40 events in total. The inversion of each simulation generates two points, one for pipe 1 (blue) and one for pipe 2 (orange). (a) Predicted mean $\Gamma^*$ versus the true assigned $\Gamma^*$ for that simulation. (b) Predicted mean $s_T^*$ versus the true assigned $s_T^*$ for that simulation. \label{fig:inv_two_uncoupled}}
\end{figure}

\begin{figure}[!htbp]
\centering
\includegraphics[width=1.0\linewidth]{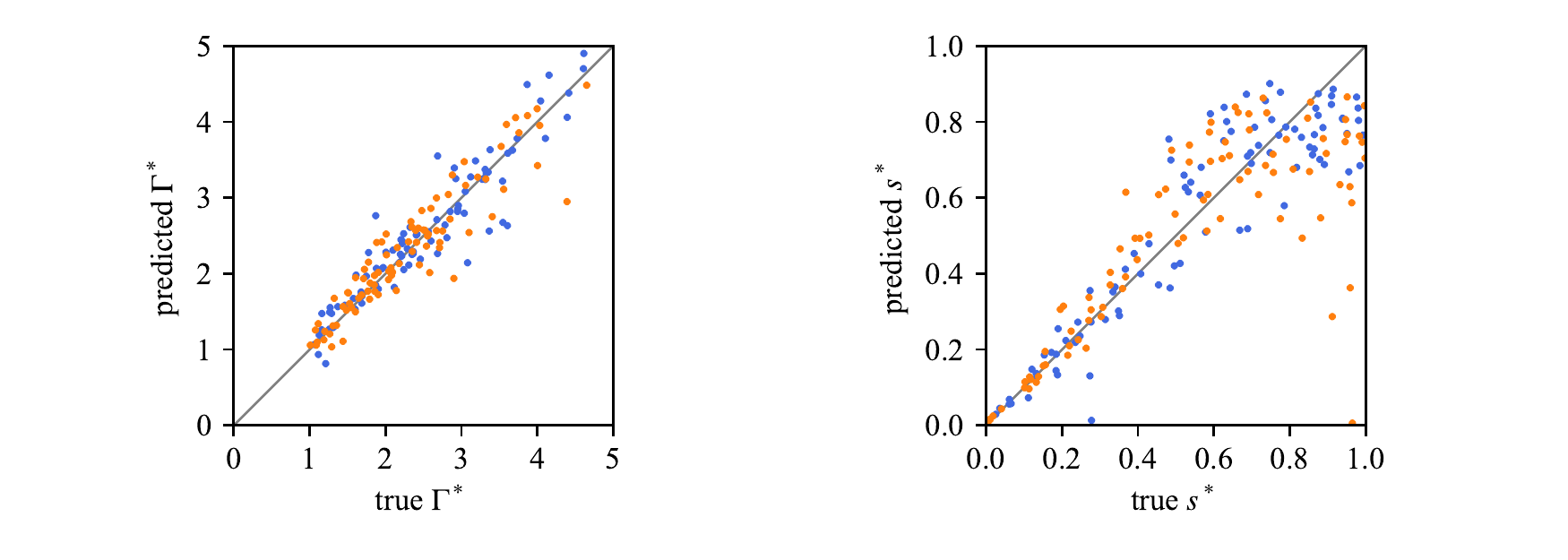}
\caption{Showing inversion results for simulated synthetic data of two coupled pipes. Each simulation uses a pair of values for $\Gamma^*$ and $s_T^*$ to generate 40 events in total.  The inversion of each simulation generates two points, one for pipe 1 (blue) and one for pipe 2 (orange). (a) Predicted mean $\Gamma^*$ versus the true assigned $\Gamma^*$ for that simulation. (b) Predicted mean $s_T^*$ versus the true assigned $s_T^*$ for that simulation. \label{fig:inv_two_coupled}}
\end{figure}

We similarly perform inversions on synthetic data from simulations of two uncoupled pipes, shown in Fig.~\ref{fig:inv_two_uncoupled}, and two coupled pipes, shown in Fig.~\ref{fig:inv_two_coupled}. In these figures, each point represents results from Bayesian inversion applied to a simulated sequence of 40 venting times from two pipes. The number of venting times was chosen to investigate the level of uncertainty in the inversion for a pair of Levant pipe trails, which each typically comprise $\sim$20~venting times. In each case, the predicted $\Gamma^*$ is in agreement with the true value; the uncertainty in the inference of $s^*$ increases with increasing true $s^*$.

\newpage
\section{Oceanus posterior distributions}
\label{sec:posterior}

\begin{figure*}[!htbp]
  \centering
  \includegraphics[width=1.0\linewidth]{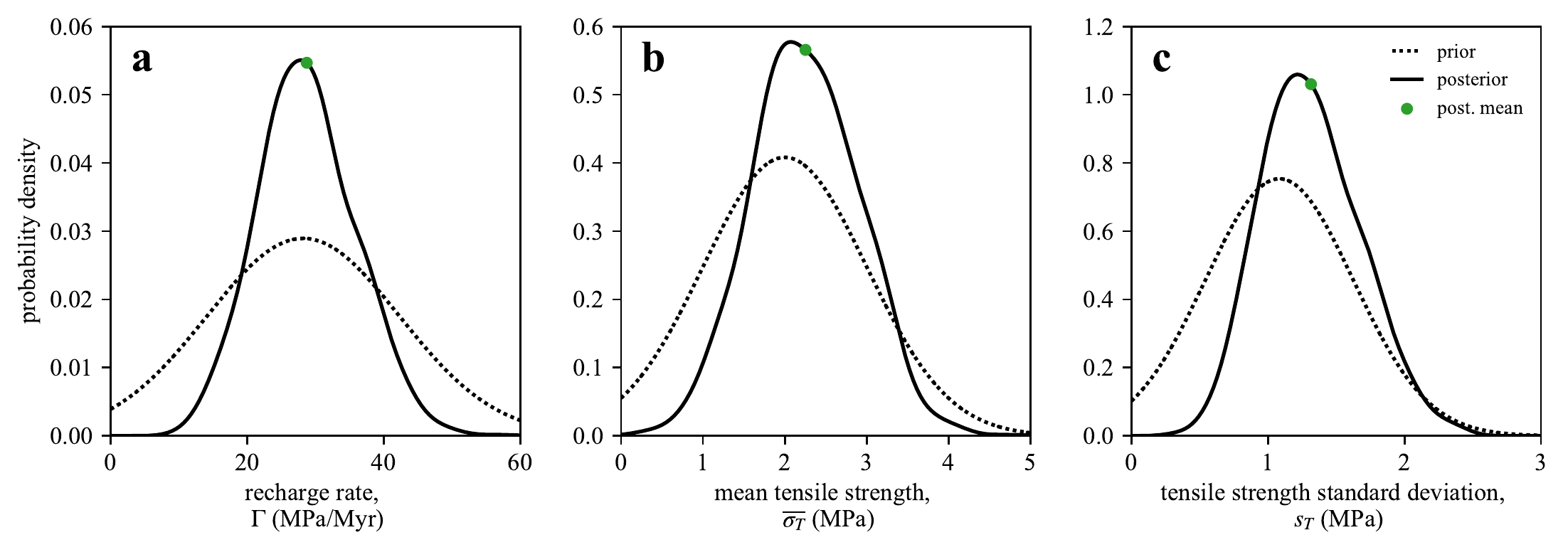}
\caption{Results of Bayesian inference applied to Oceanus trail. (a) Prior and posterior distributions of the recharge rate, $\Gamma$, with posterior mean 28~MPa/Myr. (b) Prior and posterior distributions of the mean tensile strength, $\overline{\sigma_T}$ with posterior mean 2.3~MPa. (c) Prior and posterior distributions of the tensile strength standard deviation, $s_T$ with posterior mean 1.3~MPa.}
\end{figure*}

\newpage
\section{Pressure coupling}
\label{sec:coupled}

\begin{figure}[!htbp]
\centering
\begin{subfigure}[b]{0.49\textwidth}
   \includegraphics[width=1\linewidth]{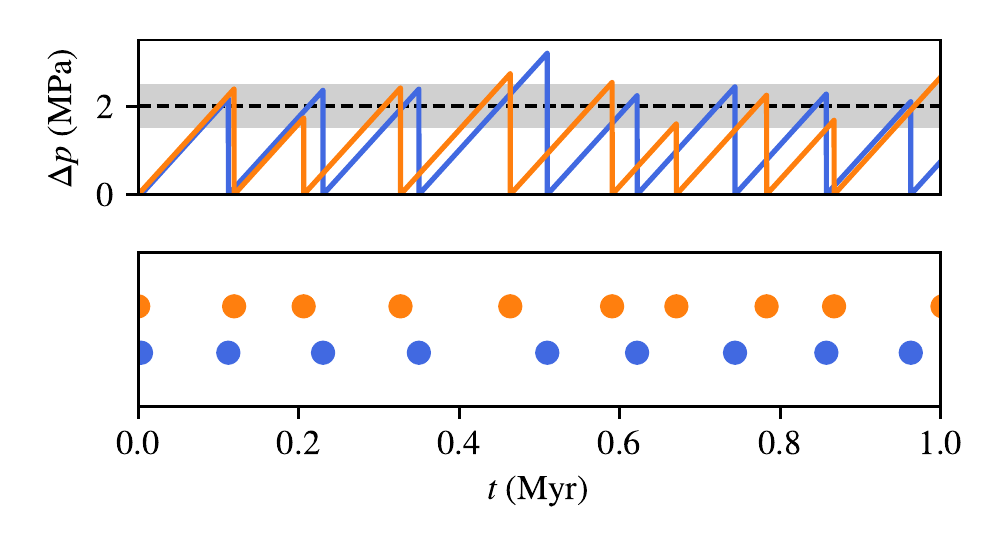}
\end{subfigure}
\begin{subfigure}[b]{0.49\textwidth}
   \includegraphics[width=1\linewidth]{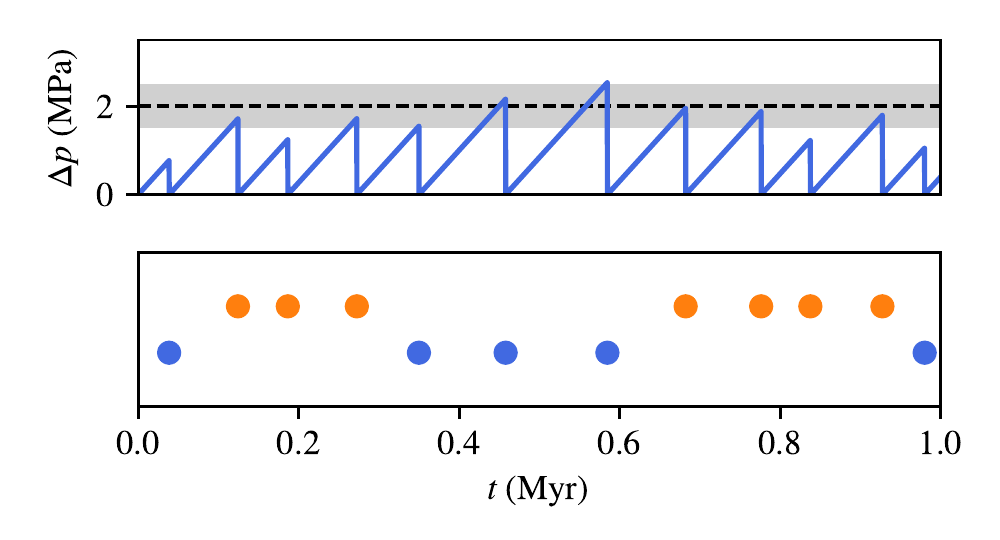}
\end{subfigure}
\caption{Stochastic realisations of two pipes, each with $\Gamma = 20$~MPa/Myr, $s = 0.5$~MPa and $\overline{\sigma_{T}} = 2$~MPa. Top plots show pressure evolution, where $\Delta p = p - \sigma_{\mathrm{min}}$ and bottom plots show the corresponding pockmark patterns. Left: uncoupled, right: coupled. Dashed horizontal lines indicate the mean tensile strength $\overline{\sigma_T}$; grey bars contain tensile strengths within one standard deviation of the mean ($\overline{\sigma_T} \pm s_T$). 
\label{fig:coupled}}
\end{figure}

Whether pipe trails are coupled or not leads to profound changes in the resulting pattern of pockmarks. Fig. \ref{fig:coupled} demonstrates this clearly with a pair of pipes in each scenario. The uncoupled pipes have independent pressure histories hence the pattern of pockmarks leads to an alternation between each pipe venting. In contrast, for a system of two coupled pipes, after either pipe vents both pressures reset to $\Delta p = 0$ and sample new tensile strengths. Therefore, if one pipe vents, the other is temporarily inhibited from venting. This leads to a complementary pockmark series, where the periods of quiescence of one pipe correspond to activity in the second pipe. In comparison to the field data (Fig. \ref{fig:1}d), uncoupled behaviour is exhibited by pipe trails 10, 11 and 12, while coupled behaviour is most pronounced in trails 3, 4 and 5.

\section{Triple-wise inference}
\label{sec:triple}

\begin{figure*}[!htbp]
  \centering
  \includegraphics[width=0.9\linewidth]{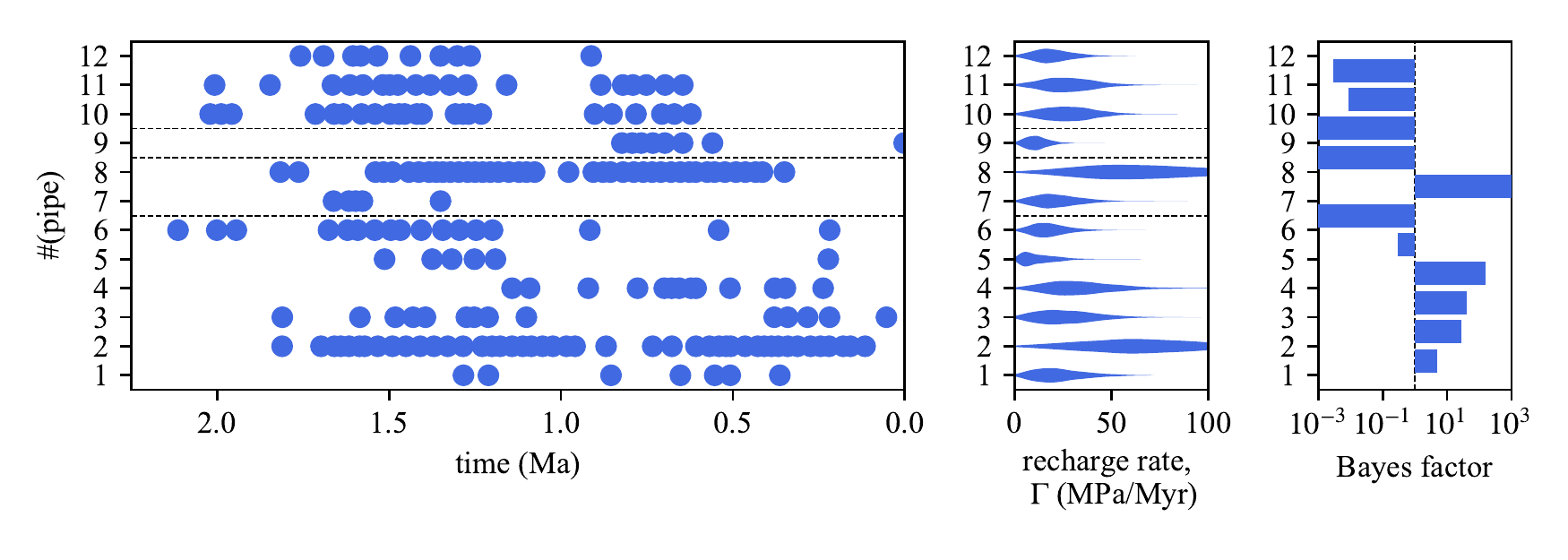}
\caption{Results of triple-wise Bayesian inference applied to Levant margin data. (a) Time-transformed data from \cite{oppo2021leaky}. Dashed lines divide pipe clusters that are separated by more than 10 km. (b) Violin plot of posterior recharge rate distributions for each pipe trail. (c) Bayes factors of pairwise pipe analysis, where a positive value implies the coupled model is more likely.}
\end{figure*}

\newpage
\section{Candidate overpressure mechanisms}
\label{sec:candidate}

\subsection{Tectonic compression}

\begin{table}[!htbp]
\centering
\begin{tabular}{lllllll}
\textbf{Parameter}              & \textbf{Description}       & \textbf{mean} & \textbf{std. dev.} & \textbf{min.} & \textbf{max.} & \textbf{Reference} \\ \hline
$\Delta P/\Delta e_{xx}$ (MPa) & overpressure per \% strain & 1.2 & 0.2 & 0.5 & 2.0 & \cite{obradors2016stress} \\ \hline
$\Delta e_{xx}$ & horizontal strain & 5 & 2 & 1 & 10 & \\ \hline
$\Delta t_e$ (Myr) & strain duration & 5.5 & 0.5 & 5 & 6 & \cite{cartwright2021quantitative}
\\[2mm]
\end{tabular}
\end{table}   

The model for tectonic compression developed by \cite{kearney2023episodic} is highly simplified; to ensure an accurate estimation of $\Gamma_s$ we use results from previous numerical modelling of tectonic compression \citep{obradors2016stress}. \cite{obradors2016stress} estimate overpressures between 11--14.2~MPa from 10\% strain at different rates, implying that 1.1--1.4~MPa of overpressure is generated per \% strain. Using seismic imaging, we estimate the strain at Oceanus to range from 1\% to 10\%, which we assume has been accumulating since the Messinian Salinity Crisis, between 5~Ma to 6~Ma. 

\subsection{Pressure diffusion}

\begin{table}[!htbp]
\centering
\begin{tabular}{lllllll}
\textbf{Parameter}              & \textbf{Description}       & \textbf{mean} & \textbf{std. dev.} & \textbf{min.} & \textbf{max.} & \textbf{Reference} \\ \hline
$h_s$ (m)                       & sandstone thickness        & 150           & 50                 & 50            & 200           & \cite{cartwright2021quantitative}, \\
$h_m$ (m)                       & mudstone thickness         & 2500          & 250                & 2000          & 3000          &                               \\ \hline
$\alpha_s$                      & sandstone Biot coefficient & 0.62          & 0.17               & 0.38          & 0.83          & \cite{ge1992hydromechanical}                              \\
$\alpha_m$                                                                                                                               & mudstone Biot coefficient  & 0.68          & 0.35               & 0.30          & 0.98          &                               \\
$v_s$           & sandstone Poisson ratio    & 0.24          & 0.04               & 0.20          & 0.30          &                               \\
$v_m$           & mudstone Poisson ratio     & 0.25          & 0.05               & 0.15          & 0.30          &                               \\
$K_s$ (GPa)                     & sandstone bulk modulus     & 18            & 8                  & 8             & 30            &                               \\
$K_m$ (GPa)                     & mudstone bulk modulus      & 15            & 17                 & 5             & 33            &                               \\ \hline
$\phi_s$                        & sandstone porosity         & 0.22          & 0.01               & 0.19          & 0.24          & \cite{ortega2018dynamic}                              \\ \hline
$\phi_m$                        & mudstone porosity         & 0.20          & 0.05               & 0.05          & 0.30          &  \cite{yang2007permeability} \\
log$_{10}$$k_m$ (log$_{10}\,$m$^2$)                       & (log) mudstone permeability         & -19          & 0.5                & -22          & -18          &                               \\ \hline
$\eta$ (mPa$\,$s)                       & water viscosity         & 0.3          & 0.1              & 0.1          & 0.5          & \cite{abramson2007viscosity} \\ \hline
$c_\ell$ (10$^{-11}$ Pa$^{-1}$) & water compressibility      & 4.0           & 0.1                & 3.7           & 4.3           & \cite{fine1973compressibility} \\[2mm]
\end{tabular}
\end{table}

\subsection{Flow focusing: marginal uplift}

\begin{table}[!htbp]
\centering
\begin{tabular}{lllllll}
\textbf{Parameter}              & \textbf{Description}       & \textbf{mean} & \textbf{std. dev.} & \textbf{min.} & \textbf{max.} & \textbf{Reference} \\ \hline
$\theta$ ($^\circ$) & tilt angle & 3 & 1 & 0 & 5 & \cite{cartwright2021quantitative} \\
$\rho_m$ (kg/m$^3$)& mudstone density & 2350 & 100 & 2200 & 2600 & \\
$\rho_\ell$  (kg/m$^3$) & water density & 1060 & 100 & 1000 & 1200 & \\
 $L_s$ (m) & sandstone length & 5000 & 1000 & 500 & 10000 & \\ \hline
$\Delta t_u$ (Myr) & uplift duration & 2 & 0.5 & 1 & 3 & \cite{oppo2021leaky}
\\[2mm]
\end{tabular}
\end{table}   

Flow focusing due to marginal uplift can lead to overpressure generation. For a flat sandstone of length $L_s$ in a mudstone with pressure gradient $\rho_m g$, uplifting one side by $\infd z$ leads to an equilibration of pressures at the new sandstone centroid, $\infd z/2$ \citep{flemings2002flow}. Therefore, tilting the sandstone by an angle $\theta$ gives an overpressure of
\begin{equation}
    \Delta p = \tfrac{1}{2} (\rho_m - \rho_\ell) g L_s \sin \theta,
\end{equation}
at the top of the reservoir. The corresponding overpressure rate is
\begin{equation}
    \pd{p}{t} = \tfrac{1}{2}(\rho_m - \rho_\ell) g L_s \dot \theta \cos \dot \theta t,
\end{equation}
where $\dot \theta$ is the angular tilting rate. For simplicity of interpretation, we take the time-average of this overpressure rate, given by
\begin{equation}
    \Gamma = \frac{(\rho_m - \rho_\ell) g L_s \sin \theta}{2 \Delta t_u} .
\end{equation}

\subsection{Flow focusing: folding}

\begin{table}[!htbp]
\centering
\begin{tabular}{lllllll}
\textbf{Parameter}              & \textbf{Description}       & \textbf{mean} & \textbf{std. dev.} & \textbf{min.} & \textbf{max.} & \textbf{Reference} \\ \hline
$\rho_m$ (kg/m$^3$) & mudstone density & 2350 & 100 & 2200 & 2600 & \cite{cartwright2021quantitative}\\
$\rho_\ell$ (kg/m$^3$) & water density & 1060 & 100 & 1000 & 1200 & \\
$h_s$ (m)                       & sandstone thickness        & 150           & 50                 & 50            & 200           & \\
$\Delta t_f$ (Myr) & folding duration & 5.5 & 0.5 & 5 & 6 & 
\\[2mm]
\end{tabular}
\end{table}   

If the sandstone reservoir has a growing parabolic profile, then the overpressure rate generated at the crest by flow focusing is given by \citep{flemings2002flow}
\begin{equation}
    \Gamma = \frac{2(\rho_m - \rho_\ell) g \Delta h_f}{3\Delta t_f}
\end{equation}
where the factor of $2/3$ appears because the sandstone and mudstone pressures equilibrate at $\Delta h/3$. 

\subsection{Disequilibrium compaction}

\begin{table}[!htbp]
\centering
\begin{tabular}{lllllll}
\textbf{Parameter}              & \textbf{Description}       & \textbf{mean} & \textbf{std. dev.} & \textbf{min.} & \textbf{max.} & \textbf{Reference} \\ \hline
$\rho_{ps}$ (kg/m$^3$) & post-salt sediment density & 2000 & 100 & 1800 & 2500 & \cite{cartwright2021quantitative} \\
$h_{ps}$ (m) & post-salt sediment thickness & 355 & 25 & 300 & 400 & \\
$\Delta t_c$ (Myr) & duration & 5 & 0.5 & 4 & 6 & 
\\[2mm]
\end{tabular}
\end{table}

Disequilibrium compaction due to post-salt sedimentation contributes to overpressure in the North Levant Basin. For a change in post-salt sediment thickness $\Delta h_{ps}$ over a time $\Delta t_c$ with density $\rho_{ps}$, the maximum overpressure rate is given by the change in total stress,
\begin{equation}
    \Gamma = \frac{\rho_{ps} g \Delta h_{ps}}{\Delta t_c}
\end{equation}

\end{document}

%% file: Other/commands.tex
\newcommand{\infd}{\text{d}}

\newcommand{\pd}[2]{\frac{\partial{#1}}{\partial{#2}}}

\newcommand{\prob}{\mathbb{P}}
\newcommand{\given}{\, | \,}

\newcommand{\history}{\mathcal{H}_{t_n}}

\newcommand*{\rom}[1]{\expandafter\@slowromancap\romannumeral #1@}